\documentclass[fleqn,10pt]{wlscirep}

\usepackage{graphicx}
\usepackage{amsmath,amssymb,bm}
\usepackage[version=3]{mhchem} 
\usepackage{subfigure}
\usepackage{siunitx}
\usepackage{dsfont}

\title{Transverse Anderson localization of light near Dirac points of photonic nanostructures}

\author[1,2]{ Hanying Deng}
\author[1,2,+]{Xianfeng Chen}
\author[3]{Boris A. Malomed}
\author[4]{Nicolae~C.~Panoiu}
\author[1,2,*]{Fangwei Ye}
\affil[1]{Department of Physics and Astronomy, Shanghai Jiao Tong University, Shanghai 200240, China}

\affil[2]{Key Laboratory for Laser Plasma (Ministry of Education), IFSA Collaborative Innovation Center, Shanghai Jiao Tong University, Shanghai 200240, China}

\affil[3]{Department of Physical Electronics, School of Electrical Engineering, Faculty of Engineering, Tel Aviv University, Tel Aviv 69978, Israel}

\affil[4]{Department of Electronic and Electrical Engineering,
University College London, Torrington Place, London WC1E7JE, UK}

\affil[*]{fangweiye@sjtu.edu.cn}
\affil[+]{xfchen@sjtu.edu.cn}

\begin{abstract}
We perform a comparative study of the Anderson localization of light beams in disordered layered photonic nanostructures that, in the limit of periodic layer distribution, possess either a Dirac point or a Bragg gap in the spectrum of the wavevectors. In particular, we demonstrate that the localization length of the Anderson modes increases when the width of the Bragg gap decreases, such that in the vanishingly small bandgap limit, namely when a Dirac point is formed, even extremely high levels of disorders are unable to localize the optical modes located near the Dirac point. A comparative analysis of the key features of the propagation of Anderson modes formed in the Bragg gap or near the Dirac point is also presented. Our findings could provide valuable guidelines in assessing the influence of structural disorder on the functionality of a broad array of optical nanodevices.
\end{abstract}
\begin{document}

\flushbottom
\maketitle
%
%
%

\section*{Introduction}

Periodic photonic structures provide a versatile platform to control and engineer light-matter interaction as the photon dispersion in such patterned optical media can be designed to differ significantly from the optical wave dispersion in regular materials\cite{PhCbook}. In particular, various exotic photonic bandstructures can be engineered, including frequency gaps, spectral domains with strong anisotropy, negative refraction, and Dirac-points (DPs). In the case of DPs, whose properties are at the center of this study, an upper and lower photonic bands are designed in such a way that they intersect at a single point, the frequency dispersion of the optical modes located in the vicinity of this point being linear. In condensed matter physics, the electronic counterpart of photonic DPs are at the origins of many remarkable properties of recently discovered materials, such as graphene and topological insulators \cite{geim2007rise, novoselov2005two, rechtsman2013photonic, orlita2008approaching,xia2009observation}.  Therefore, the ability to create, eliminate, and more generally manipulate DPs of photonic structures could have many important implications, both at basic science level as well as to photonic devices with new or improved functionality. For example, photonic DPs provide an alternative, more convenient way to explore and understand DP-related physics, as photonic structures offer a versatile, easy to use platform for the experimental implementation of such physical
systems \cite{topolancik2007experimental, bayindir2001photonic}.

 Generally speaking, photonic DPs can be divided into two broad classes according to how they form. The first class encompasses structural DPs \cite{peleg2007conical, ochiai2009photonic, zhuyi2009,diem2010transmission, kartashov2013light, moti2013}, the existence of these DPs being related to specific topological and structural properties of the corresponding photonic lattice. In particular, these DPs are formed in honeycomb lattices regardless of the lattice parameters and can be viewed as the photonic analog of the electronic DPs of graphene and other two-dimensional (2D) materials. The second class of DPs includes the so-called
accidental-degeneracy-induced DPs (ADIDPs) \cite{zsn2009,huang2011dirac, chen2014accidental, sakoda2014photonic}. These DPs can appear in simple lattices such as square lattices when the lattice parameters (lattice constant, refractive index contrast) are finely tuned in such a way that at specific frequencies the effective permittivity and permeability of the structure
are zero \cite{huang2011dirac}. Interestingly, it has been recently demonstrated that the band structure of surface-plasmon polaritons (SPPs) formed in one-dimensional (1D) layered metallic-dielectric nanostructures
possesses ADIDPs \cite{nam2010diabolical, sun2013giant}, provided that the spatial average of the permittivity of the lattice is zero. By incorporating graphene layers into such periodic nanostructures, electrically or optically tunable DPs have been shown as well \cite{deng2015optically}.

In reality, however, fabrication processes inherently introduce a certain level of structural disorder upon producing periodic structures, and thus it is important to investigate whether this disorder would destroy the DPs thus preventing the experimental observations of DP-related physics. The effects of disorder on the structural DPs have been investigated in the particular case of optical honeycomb waveguide lattices, the main finding being that the DP-associated chiral symmetry would be preserved or damaged, depending on the nature of the disorder \cite{zeuner2014edge}. However, the influence of disorder on the ADIDPs, which is the aim of the present report, has not been studied yet.

In this report, we perform a comparative study of the effect of structural disorder on optical properties of one-dimensional (1D) photonic nanostructures that, in the momentum space, possess either a DP or a Bragg gap (BG). We find that DPs are extremely robust against structural disorder, despite the fact that they are formed when an accidental degeneracy occurs. More specifically, the photonic modes of the disordered lattices located near DPs of the unperturbed ones remain delocalized even when the strength of disorder is as high as \SI{80}{\percent}. The photonic modes of nanostructures having BGs, on the other hand, are much more prone to mode localization. We have established these conclusions by performing both a detailed mode analysis and direct beam propagation simulations.

\section*{Results and Discussion}

\begin{figure}[ht]
\centering
\includegraphics[width=0.5\textwidth]{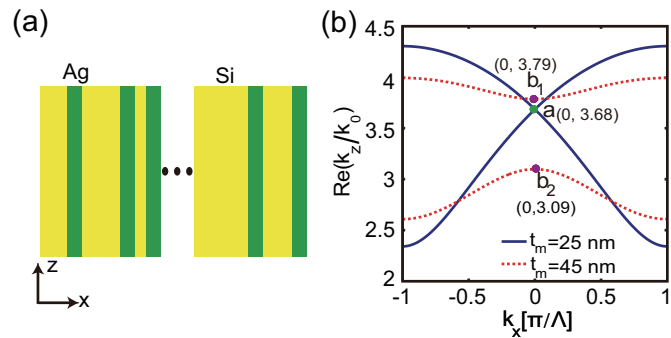}
\caption{(\textbf{a}) Schematic of the binary (metallic-dielectric) layered nanosctructure. (\textbf{b}) The photonic bandstructure of two periodic lattices, one having the thickness of the metallic layer, $t_{m}=$~\SI{25}{\nano\metre}(blue/solid lines), and the other one with thickness, $t_{m}=$~\SI{45}{\nano\metre} (red/dashed lines). A Dirac point forms in the former and BG forms in the latter. The thickness of the silicon layer is $t_{d}=$~\SI{256}{\nano\metre} in both cases. Point ``a"  denotes the location of the Dirac point; points ``$b_1$"  and ``$b_2$" denote the boundaries of the BG.} \label{fig:geometry}
\end{figure}

The photonic lattice considered in this study is a binary periodic nanostructure composed of alternating layers of metallic and dielectric materials stacked along the $x$-axis, as depicted in Figure~\ref{fig:geometry}(a). The electromagnetic field is assumed to propagate along the $z$-axis. To make our analysis more specific, we assume that the metallic and dielectric layers are made of silver and silicon, respectively. The permittivity of dielectric (silicon) is $\varepsilon_{d}=12.25$.  The complex permittivity of the metal (silver) is $\varepsilon_{m}=-125.39+2.84i$ at the wavelength of $\lambda=$~\SI{1550}{\nano\meter} \cite{johnson1972optical}. Note that throughout the Report, we have taken into account the metallic losses in the analysis. The photonic bands of the photonic lattice are obtained by using the transfer-matrix method (see Methods) and are given by the solutions of the following transcendental equation:

 \begin{eqnarray}
\label{eq:band} \cos(k_x \Lambda)=\cos(k_dt_d)\cos(k_m t_m)-\frac{1}{2}(\frac{q_d}{q_m}+\frac{q_m}{q_d})\sin(k_dt_d)\sin(k_m t_m),
\end{eqnarray}

\noindent where $k_{x}$ is the Bloch wave vector, $k_{z}$ the propagation wavevector of the Bloch wave, $%
t_{d}$, $t_{m}$ are the thickness of dielectric and metallic layers, and $\Lambda
=t_{d}+t_{m}$ is the period of the unit cell.  The wavevectors $k_{j}=\sqrt{(\omega/c)^{2}\varepsilon _{j}-k_{z}^{2}}$,
 where $q_j=\left[\mu_j/\varepsilon_j(1-\sin^{2}\theta/\mu_j\varepsilon_j)\right]^{1/2}$, ($j=d,m$), $\theta$ is the incident angle and \emph{c} is the speed of  light in the vacuum. By fixing the operating frequency,  $\omega$, in  Eq.~(\ref{eq:band}), one obtains the dependence of $k_z=k_z(k_x)$, which determines the spatial modal dispersion relation for the propagating modes at the particular frequency. The transmission bands for two different values of the thickness, $t_{m}$, are shown in Figure~\ref{fig:geometry} (b).

 It has been reported that 1D DPs forms at the center of the Brillouin zone \cite{deng2015optically}, $k_{x}=0$, provided that the real part of the spatially averaged permittivity is equal to zero, namely $\mathfrak{Re}(\overline{\varepsilon })=\mathfrak{Re}\left(\frac{\varepsilon_dt _d+\varepsilon_mt _m}{t_d+t_m}\right)=0$. As shown in Figure~\ref{fig:geometry} (b),
for $t_{m}=$~\SI{25}{\nano\metre}, the condition $\mathfrak{Re}(\overline{\varepsilon })=0$  holds, and indeed the two transmission bands touch in a single point, giving rise to a photonic DP at the position $(k_x, \mathfrak{Re}{(k_z)})=(0, 3.68k_0)$. For such a two-band configuration to occur, the thickness of the dielectric and metallic layers should be larger than certain critical values (for example, for $t_{m}=$~\SI{25}{\nano\metre}, in order to support the two-band structure, the minimum thickness of dielectric layers is \SI{199}{\nano\metre}).
This kind of DP is of ADIDP nature, as it is removed once the averaged permittivity deviates from zero. Specifically, a BG opens when the averaged permittivity is shifted away from the zero value. For example, when $t_{m}=$~\SI{45}{\nano\metre}, the photonic band structure of the binary lattice possesses a BG in the region where the DP existed (see the red curve in Figure~\ref{fig:geometry}(b) ).

 We then gradually introduce disorder into the nanostructure by assuming a random fluctuation of the thickness of the dielectric components, namely, the thickness of the $n$-th silicon layer is set to  $t_{d}^{n}=t_{d0}+\delta _{n}$, where $t_{d0}$  is the average thickness (we take $t_{d0}=$~\SI{256}{\nano\metre} here), and $\delta _{n}$ is a random value. We assume $\delta _{n}$ to be uniformly distributed in the interval of  $[-\delta ,\delta ]$, $0<\delta<t_{d0}$. hence the level of disorder can be characterized by the parameter, $\Delta \equiv \delta /t_{d0}$.

 \begin{figure}[t]
\centering \subfigure{\includegraphics[width=0.45\textwidth]{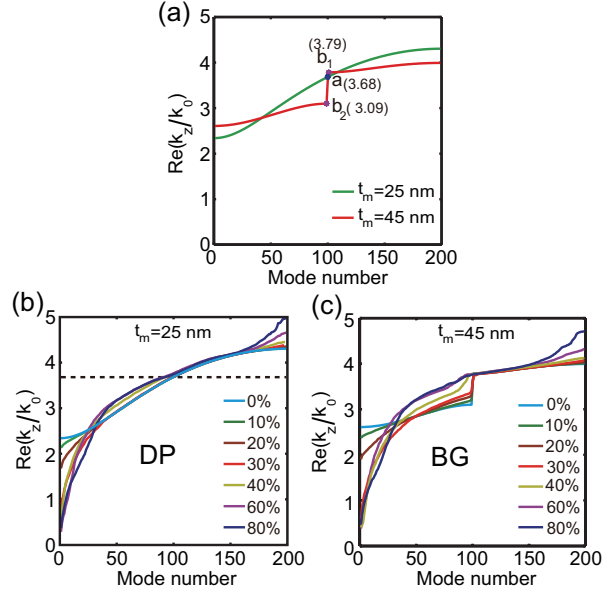}}\hspace{0.0001\textwidth}
\caption{\label{fig:mode} (\textbf{a}) Eigenvalue spectra of the Bloch modes for a periodic lattice with $t_{m}=$~\SI{25}{\nano\metre} (green line) and $t_{m}=$~\SI{45}{\nano\metre} (brown line). The influence of disorder on the eigenvalue spectra of the two lattices are shown in (\textbf{b}) and (\textbf{c}), respectively. The results in (\textbf{b}) and (\textbf{c}) are obtained by ensemble-average over \SI{50} disorder realizations.}
\end{figure}

\noindent\textbf{Effect of the disorder on the eigenvalues of Anderson modes.}  We first compare the influence of disorder on the eigenvalues of the Anderson modes of the two distinct lattices, namely the DP-bearing lattice and the lattice that possesses a BG.
In Figure~\ref{fig:mode}(a), we show the eigenvalue spectrum of these two periodic structures, one possessing a
DP (at $t_{m}=$~\SI{25}{\nano\metre}), whereas the other one being a regular lattice featuring a BG (at $t_{m}=$~\SI{45}{\nano\metre}). In the plots, the modes are indexed according to the magnitude of their associated propagation constant (mode eigenvalue).

The modification of the eigenvalue spectra when disorder is gradually introduced in the two lattices is shown in
Figures~\ref{fig:mode}(b) and \ref{fig:mode}(c), where the results are averaged over 50 randomly-perturbed configurations. In the case of the DP-bearing lattice, the eigenvalues of the eigenmodes at the top and bottom of the photonic bands experience significant shift, whereas the eigenvalues near the center of the band are only slightly affected by disorder. Remarkably, as Figure~\ref{fig:mode}(b) illustrates, even if the disorder strength is increased to $\Delta=$~\SI{80}{\percent},  the relative variation of the eigenvalues located near the DP of the periodic lattice is extremely weak (less than
~\SI{2}{\percent}).  In contrast, in the lattice that possesses a BG in its unperturbed limit, the variation of the eigenvalues upon the introduction of disorder shows a very different dynamics. Thus, the modes most affected by disorder are located at the bottom of the lower band and the larger the eigenvalues are the smaller their variation is. The eigenvalues belonging to the upper band, on the other hand, are hardly affected by disorder. Moreover, the corresponding bandgap gradually shrinks until it completely vanishes if the disorder strength is increased beyond a certain value. Of course, in such systems having a BG in their unperturbed limit, the vanishing of the bandgap at the strong disorder level does not imply the appearance of a disorder-induced DP, as the inequality, $\overline{\varepsilon } \neq0$,  remains unchanged for such systems. With a further increases of disorder, the upper band starts to be perturbed, too.

\noindent\textbf{The effective width of Anderson modes.} After gaining these valuable insights into the relations between the spectra of the Anderson modes, the topology of the band structure of the lattice, and the disorder strength, we look into the localization length of the Anderson modes. For this, we calculated the effective width of these modes by using the following definition,

\begin{eqnarray}
\label{eq:width} \mathbf{W}_{\text{eff}}=\bigg{\langle}\bigg{|}\frac{\int_{-\infty }^{+\infty }|\mathbf{E}%
(x)|^{2}(x-x_c)^{2}dx}{\int_{-\infty }^{+\infty }|\mathbf{E}%
(x)|^{2}xdx}\bigg{|}^{\frac{1}{2}}\bigg{\rangle},
\end{eqnarray}

\noindent where $x_{c}\equiv \int_{-\infty }^{+\infty }|\mathbf{E}(x)|^{2}xdx/\int_{-\infty }^{+\infty }|\mathbf{E}(x)|^{2}dx$
 is the center of the mode and $``\langle \rangle" $ stands for ensemble averaging over multiple realizations of randomness with the same level of disorder.

\begin{figure}[t]
\centering \subfigure{\includegraphics[width=0.6\textwidth]{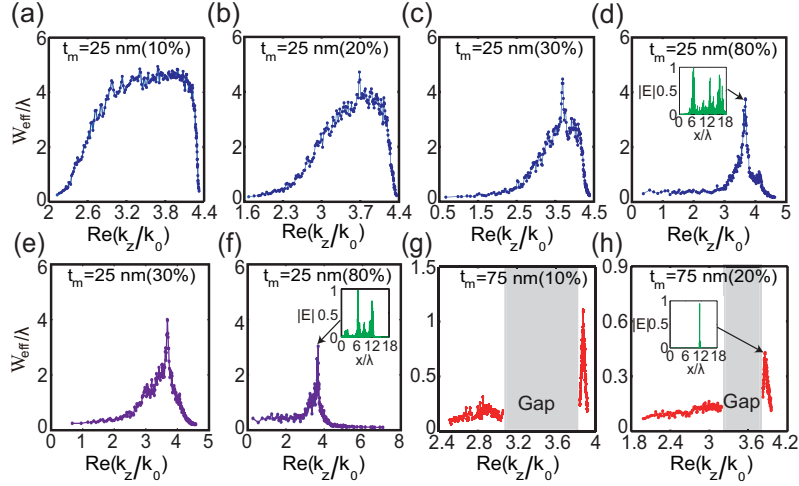}}\hspace{0.0001\textwidth}
\caption{\label{fig:disorder} The dependence of the width of the Anderson modes on the disorder level, calculated for the DP-bearing system (\textbf{a})-(\textbf{f}), and for the BG system (\textbf{g})-(\textbf{h}). In (\textbf{e}) and (\textbf{f}), both dielectric and metallic layers are assumed to be disordered. Insets in (\textbf{d}), (\textbf{f}) and (\textbf{h}) show the profiles of the most delocalized Anderson modes. The gray region in (\textbf{g}) and (\textbf{h}) indicate the bandgap of the eigenvalue spectra of the unperturbed lattices.}
\end{figure}

Figure~\ref{fig:disorder} shows the effective width of the modes of the two structures, which contain a gradually increasing degree of structural disorder. The plots correspond to the DP-bearing (Figures~\ref{fig:disorder}(a)-\ref{fig:disorder}(d)) and BG-bearing (Figures~\ref{fig:disorder}(g) and \ref{fig:disorder}(h)) lattices.  Consistent with the findings revealed by Figure~\ref{fig:mode}(b), Figures~\ref{fig:disorder}(a)-\ref{fig:disorder}(d) show that the modes with the smallest and largest eigenvalues firstly become localized, then the mode localization effect extends towards the central region of the spectrum. Remarkably, however, the modes at the central region of the spectrum, which correspond to the DP of the unperturbed lattice, remain delocalized despite the fact that the disorder level is increased to a particularly large value (\SI{80}{\percent} in  Figure~\ref{fig:disorder}(d)). As expected, a similar scenario is as well observed when the thickness of the dielectric and metallic components of the unit cell of the lattice are both perturbed (Figures~\ref{fig:disorder}(e) and \ref{fig:disorder}(f)).
 The spatial profile of the mode with the largest effective width is also shown in the inset of Figure~\ref{fig:disorder}(d) and \ref{fig:disorder} (f). Interestingly enough, we find that the widest mode has a propagation constant that is exactly equal to the propagation constant of the corresponding DP. This is a clear manifestation of an extreme robustness of such ADIDP against structural disorder, a phenomenon that can be understood by recalling that the condition for the formation of such ADIDPs is that the averaged permittivity is zero, namely, $\bar{\varepsilon}=0$. Thus, in the unperturbed limit, if one has
$\overline{\varepsilon }=\frac{\varepsilon_dt _{d0}+\varepsilon_mt _{m0}}{t_{d0}+t_{m0}}=0$, then in the disordered lattice the
averaged permittivity remains zero as,

 \begin{eqnarray}
\bar{\varepsilon}=\lim\limits_{\mathbf{L} \rightarrow \infty} \frac{\sum\limits_{n}\varepsilon_m(t_{m0}+\delta^{n} _{m})+\sum\limits_{n}\varepsilon_d(t_{d0}+\delta^{n} _{d}) }{\mathbf{L}}=\lim\limits_{\mathbf{L} \rightarrow \infty} \frac{\sum\limits_{n}(\varepsilon_mt_{m0}+\varepsilon_dt_{d0})+\varepsilon_m\sum\limits_{n}\delta^{n} _{m}+\varepsilon_d\sum\limits_{n}\delta^{n} _{d}}{\mathbf{L}}=0 .
\end{eqnarray}
Therefore, random structural fluctuations preserve the zero-epsilon condition and consequently the corresponding mode remains delocalized.
By contrast, the width of the modes of BG-bearing lattice rapidly become localized as disorder is added to the photonic lattice. This is illustrated in Figures~\ref{fig:disorder}(g)-\ref{fig:disorder}(h), where the width of the modes corresponding to two different BG-bearing lattices are presented. In particular, one can see that even a small degree of disorder ($\Delta=$~\SI{20}{\percent})  can localize almost all modes. Note that even the mode with the largest width is tightly localized when the disorder level is
a mere \SI{20}{\percent}, as per the inset of Figure~\ref{fig:disorder}(f); compare also the widths of the most delocalized modes of the two lattices, shown in the insets of Figures~\ref{fig:disorder}(d), \ref{fig:disorder}(f) and \ref{fig:disorder}(h).

\begin{figure}[t]
\centering \subfigure{\includegraphics[width=0.45\textwidth]{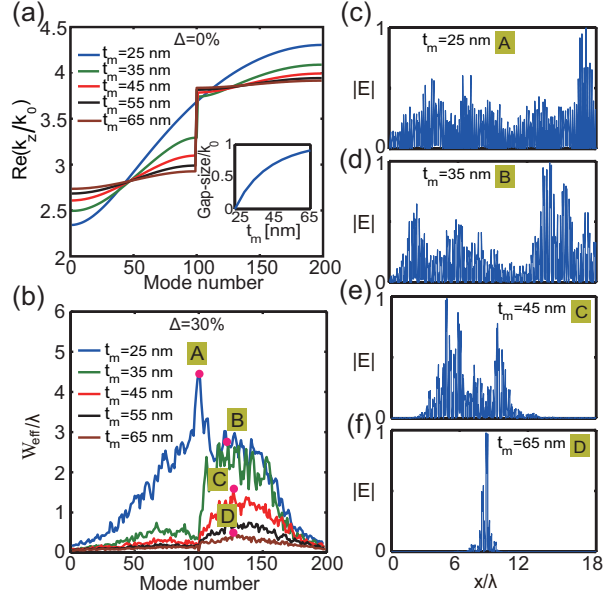}}\hspace{0.0001\textwidth}
\caption{\label{fig:bandsize} The influence of the size of the bandgap on the localization of Anderson modes. (\textbf{a}) Eigenvalue spectrums of the ideal periodic lattice with $t_{m}=$~\SI{25}, \SI{35}, \SI{45}, \SI{55}, \SI{65}{\nano\metre}. Inset: The variation of the bandgap-size with $t_{m}$. (\textbf{b}) The effective width of Anderson modes when the disorder level is \SI{30}{\percent}, determined for different values of the thickness $t_{m}$. (\textbf{c})-(\textbf{f}) The profiles of the broadest Anderson
modes.}
\end{figure}

We next investigate the relation between the bandgap-size and the degree of localization of Anderson modes. To this end, we show in Figure~\ref{fig:bandsize}(a) the eigenvalue spectra of a periodic lattice with $t_{m}=$~\SI{25}, \SI{35}, \SI{45}, \SI{55}, \SI{65}{\nano\metre}. As mentioned above, $t_{m}=$~\SI{25}{\nano\metre} corresponds to a DP-bearing system with vanishing bandgap,
and a BG with increasing width opens as the thickness of the metal $t_{m}$ increases, starting from $t_{m}=$~\SI{25}{\nano\metre}.
In Figure~\ref{fig:bandsize}(b) we present the effective widths of the eigenmodes of the corresponding perturbed lattices, the disorder parameter being the same, $\Delta=$~\SI{30}{\percent}.  Note that since the thickness of the metal layers varies across these lattices, in their unperturbed limit they possess bandgaps with different sizes. The widest eigenmodes for these different lattices are presented in Figures~\ref{fig:bandsize}(c)-\ref{fig:bandsize}(f). These figures show that a wider bandgap leads to a stronger mode localization, which implies that, at the same strength of disorder, extended Bloch modes transform into more localized Anderson modes in photonic lattices with larger bandgap-size.

\begin{figure}[t]
\centering \subfigure{\includegraphics[width=0.45\textwidth]{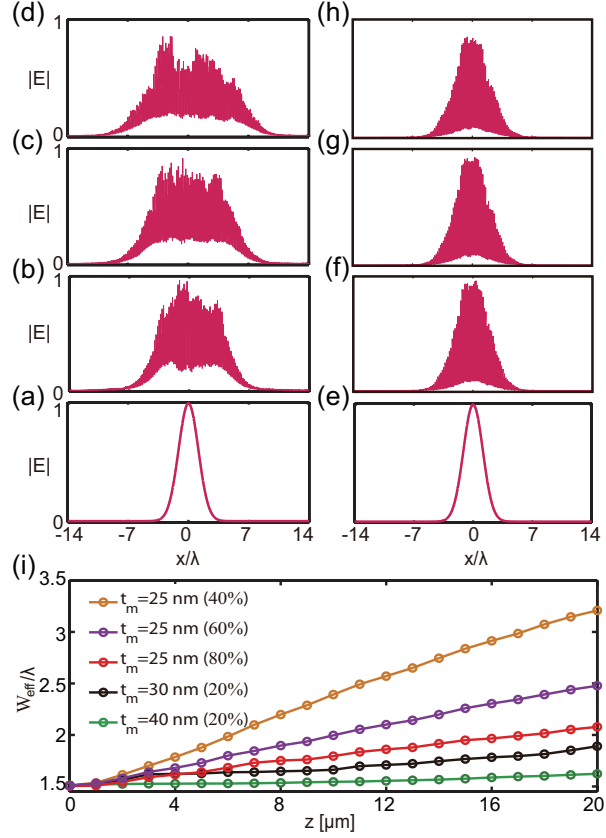}}\hspace{0.0001\textwidth}
\caption{\label{fig:propagation} The simulation of beam propagation for the DP-bearing system (\textbf{a})-(\textbf{d}) and BG system (\textbf{e})-(\textbf{h}). The disorder level is \SI{40}{\percent} in the DP-bearing system (\textbf{a})-(\textbf{d}) and \SI{20}{\percent} in the BG system (\textbf{e})-(\textbf{h}). From the bottom panel to the top one, the propagation distance is \SI{0}{\micro\meter}, \SI{10}{\micro\meter}, \SI{15}{\micro\meter}, \SI{20}{\micro\meter}, respectively. (\textbf{i}) Evolution of the effective width of the light beam during propagation.}
\end{figure}

\noindent\textbf{Simulation of the light beam propagation in disordered lattices.} Direct numerical simulations of light beams propagating in the disordered lattices investigated in this work, performed by solving numerically the 3D Maxwell equations governing the beams dynamics, corroborate the findings reported above regarding the mode properties. Figure~\ref{fig:propagation}
illustrates how an input TM-polarized Gaussian beam evolves in the lattice that in its unperturbed limit possesses a DP  (Figures~\ref{fig:propagation}(a)-(d) and in BG-bearing lattice (Figures~\ref{fig:propagation}(e)-(h)). Thus, a light beam continuously broadens in the transverse direction upon its propagation in the DP-bearing structures, despite the fact that large disorder is added into the lattice (\SI{40}{\percent} in  Figure~\ref{fig:propagation}, left panels). By contrast, in the BG-bearing lattice, a relatively much weaker disorder level could already arrest the beam transverse diffraction, leading to Anderson localization ( Figure~\ref{fig:propagation}, right panels).

The variation of the effective width of the light beam vs. propagation distance in three lattices characterized by different size of the bandgap is shown in Figure~\ref{fig:propagation}(i). The level of disorder introduced in the lattices is \SI{40}{\percent}, \SI{60}{\percent} and \SI{80}{\percent}, respectively, for $t_{m}=$~\SI{25}{\nano\metre} (DP-bearing structure),  as well as \SI{20}{\percent} for the structures with $t_{m}=$~\SI{30}{\nano\metre} and $t_{m}=$~\SI{40}{\nano\metre}.  It can be clearly seen in this figure that, despite significant disorder, the light beams propagating in DP-bearing system ($t_{m}=$~\SI{25}{\nano\metre}) undergo a much stronger wave diffraction. By contrast, even when a much weaker disorder is considered, BG-bearing
systems ($t_{m}=$~\SI{30}, \SI{40}{\nano\metre}) display a significantly reduced beam diffraction, the beam broadening being nearly halted for the structure with $t_{m}=$~\SI{40}{\nano\metre}.

\section*{Conclusions}
In summary, using both the mode analysis and direct beam propagation simulations, we have studied the influences of the structural disorder on properties of the Dirac point in one-dimensional metallic-dielectric nanostructures, and compared it with that of the Bragg-gap-bearing nanostructures.  We have demonstrated that the Dirac point is extremely robust against disorder added to the systems, in the sense that even a very large level of disorder (\SI{80}{\percent})  is unable to localize modes residing near the Dirac point. The dependence of the localization of Anderson modes on the size of the bandgap has also been explored, and it has been found that photonic modes of lattices with increasing bandgap-size in the unperturbed limit are more prone to disorder-induced localization. The extreme robustness of Dirac points found in this report points out the feasibility of the experimental observations of the rich DP-related physics. On the other hand, however, our findings points out that localizing modes near the Dirac point of the photonic lattice becomes even more critical, in contrast to the localization of the Dirac fermions in the graphene sheet\cite{morozov2006strong,tikhonenko2008weak, tikhonenko2009transition, trambly2010localization}. Finally, we should mention that,
while the results reported here were derived in the case of one-dimensional photonic structures, the finding of extreme robustness of Dirac points applies to higher-dimensional structures as well because the existence condition for the Dirac points, namely, the average permittivity is zero, is independent on system dimensionality.

\section*{Methods}

\noindent\textbf{The transfer matrix method (TMM).} We use the transfer matrix method (TMM) \cite{yeh1977electromagnetic} to obtain the photonic bandstructure of the ideal periodic photonic lattice Eq.~(\ref{eq:band}) Thus, the electromagnetic field at two positions $x$ and $x+\Delta x$ in the same layer is related via a transfer matrix \cite{nam2010diabolical, sun2013giant}:

\begin{equation}
\label{eq:TMM1} M_j(\Delta x,\omega)=\left(
  \begin{array}{cc}
   \cos(k_j\triangle x) & \frac{\displaystyle i}{\displaystyle q_j} \sin(k_j\triangle x) \\
    iq_j \sin(k_j\triangle x) &\cos(k_j\triangle x) \\
  \end{array}
\right),
\end{equation}

where $k_j=\sqrt{(\omega/c)^{2}\varepsilon_j\mu_j-k_z^{2}}$, $j = d, m$. Here, $d$ and
$m$ stand for the silicon (dielectric) and silver (metal), respectively. For a TM-polarized wave,
$q_j=\left[\mu_j/\varepsilon_j(1-\sin^{2}\theta/\mu_j\varepsilon_j)\right]^{1/2}$.

   According to the Bloch theorem, the electric and magnetic components of an electromagnetic mode in
layer $N$ and $N+1$, with Bloch wavevector, $k_x$, are related to each other through,

\begin{equation}
 \label{eq:TMM2} \left(
  \begin{array}{cc}
   E_N \\ H_N \\
  \end{array}
\right)=e^{ik_x\Lambda} \left(
  \begin{array}{cc}
     E_{N-1} \\ H_{N-1} \\
  \end{array}
\right)
\end{equation}

On the other hand, the TMM leads to:
\begin{align}
 \label{eq:TMM3}  \left(\begin{array}{cc}E_N \\ H_N  \\\end{array}\right)=M_dM_m\left( \begin{array}{cc} E_{N-1} \\ H_{N-1} \\\end{array}\right)\equiv\left(\begin{array}{cc}m_{11}& m_{12} \\m_{21} & m_{22} \\ \end{array}\right)\left(
  \begin{array}{cc}E_{N-1} \\ H_{N-1} \\ \end{array}\right).
\end{align}

Combining Eq.~(\ref{eq:TMM2})and Eq.~(\ref{eq:TMM3}) yields:

\begin{equation}
 \label{eq:TMM4} \left(
  \begin{array}{cc}
   m_{11}& m_{12} \\m_{21} & m_{22} \\
  \end{array}
\right)\left(
  \begin{array}{cc}
     E_{N-1} \\ H_{N-1} \\
  \end{array}
\right)=e^{ik_x\Lambda}\left(
  \begin{array}{cc}
     E_{N-1} \\ H_{N-1} \\
  \end{array}
\right),
\end{equation}

so that the factor $e^{ik_x\Lambda}$ is the eigenvalue of the transfer matrix $M=M_dM_m$. This
conclusion can be expressed as
\begin{equation}
 \label{eq:TMM5} e^{ik_x\Lambda}=\frac{1}{2}(m_{11}+m_{22})\pm\left[\frac{1}{4}(m_{11}+m_{22})^{2}-1\right]^{\frac{1}{2}}.
\end{equation}

From this equation one can easily derive the dispersion relation, $k_x=k_x(k_z,\omega)$,
\begin{equation}
 \label{eq:TMM6} k_x(k_z,\omega)=\frac{1}{\Lambda}\arccos\left[\frac{1}{2}(m_{11}+m_{22})\right].
\end{equation}

By substituting the elements of the matrix $M$ in Eq.~(\ref{eq:TMM6}) , this relation yields,

 \begin{eqnarray}
\label{eq:TMM7} \cos(k_x \Lambda)=\cos(k_dt_d)\cos(k_m t_m)-\frac{1}{2}(\frac{q_d}{q_m}+\frac{q_m}{q_d})\sin(k_dt_d)\sin(k_m t_m),
\end{eqnarray}

\noindent\textbf{Mode analysis and beam propagation.} The eigenmodes of the disordered photonic lattices and the optical beam propagation are numerically investigated by solving the 3D Maxwell equations using the finite element method, as implemented in COMSOL Multiphysics 4.4. The mode solver of COMSOL is used to find the eigenmodes, with the simulation domain being surrounded by scattering boundary conditions (SBCs). The frequency domain solver of COMSOL is used to simulate the optical beam propagation. A TM-polarized Gaussian beam with an x-component of electric filed, $E_{x}(x)=\exp (-x^{2}/(3\lambda)^{2})$, is used as the profile of the optical beam at the input facet of the optical superlattice. Appropriate SBCs were used to emulate open boundaries.


\section*{Acknowledgements}

The work of H.D. and F.Y. was supported by Innovation Program of Shanghai Municipal Education Commission (Grant No. 13ZZ022) and the National Natural Science Foundation of China (Grants No. 11104181 and No. 61475101). N.C.P. acknowledges financial support from the European Research Council/ERC Grant Agreement No. ERC-2014-CoG-648328.

\section*{Correspondence}
Correspondence and requests for materials should be addressed to Fangwei Ye (email:
fangweiye@sjtu.edu.cn).

\section*{Author contributions statement}
All authors have made substantial intellectual contributions to the research work. F.Y. designed the idea of the research. H. D. and F. Y. performed the numerical simulations. X. C. supervised the numerical simulations work. N.C.P. and B.A.M., together with the other authors, provided advice and performed the theoretical analysis of the results. All authors contributed to the writing of the manuscript.

\section*{Additional information}

Competing financial interests: The authors declare no competing financial interests.
\end{document}